\documentclass[12pt,preprint]{emulateapj}

\newcommand{\beq}{\begin{equation}}
\newcommand{\eeq}{\end{equation}}
\newcommand{\bdi}{\begin{displaymath}}
\newcommand{\edi}{\end{displaymath}}

\newcommand{\rmicron}{$\,\mu$m}
\def\lsim{\,\lower2truept\hbox{${<\atop\hbox{\raise4truept\hbox{$\sim$}}}$}\,}
\def\gsim{\,\lower2truept\hbox{${>\atop\hbox{\raise4truept\hbox{$\sim$}}}$}\,}

\shorttitle{BLAST: Correlations in the Cosmic Far-Infrared Background}
\shortauthors{Viero, M.~P.~et al.}

\begin{document}

\title{BLAST: Correlations in the Cosmic Far-Infrared Background at 250, 350, and 500 \rmicron\ Reveal Clustering of Star-Forming Galaxies }

\author{ Marco~P.~Viero,\altaffilmark{1, \dag}
        Peter~A.~R.~Ade,\altaffilmark{2}
        James~J.~Bock,\altaffilmark{3}
        Edward~L.~Chapin,\altaffilmark{4}
        Mark~J.~Devlin,\altaffilmark{5}
        Matthew~Griffin,\altaffilmark{2}
        Joshua~O.~Gundersen,\altaffilmark{6}
        Mark~Halpern,\altaffilmark{4}
        Peter~C.~Hargrave,\altaffilmark{2}
        David~H.~Hughes,\altaffilmark{7}
        Jeff~Klein,\altaffilmark{5}
        Carrie~J.~MacTavish,\altaffilmark{8}
        Gaelen~Marsden,\altaffilmark{4}
        Peter~G.~Martin,\altaffilmark{9,1}
        Philip~Mauskopf,\altaffilmark{2}
        Lorenzo~Moncelsi,\altaffilmark{2}
        Mattia~Negrello,\altaffilmark{10}
        Calvin~B.~Netterfield,\altaffilmark{1,11}
        Luca~Olmi,\altaffilmark{12,13}
        Enzo~Pascale,\altaffilmark{2}
        Guillaume~Patanchon,\altaffilmark{14}
        Marie~Rex,\altaffilmark{5}
        Douglas~Scott,\altaffilmark{4}
        Christopher~Semisch,\altaffilmark{5}
        Nicholas~Thomas,\altaffilmark{6}
        Matthew~D.~P.~Truch,\altaffilmark{5}
        Carole~Tucker,\altaffilmark{2}
        Gregory~S.~Tucker,\altaffilmark{15}
        Donald~V.~Wiebe\altaffilmark{4,11}}

\altaffiltext{1}{Department of Astronomy \& Astrophysics, University of Toronto, 50 St. George Street, Toronto, ON\ M5S~3H4, Canada}

\altaffiltext{2}{Department of Physics \& Astronomy, Cardiff University, 5 The Parade, Cardiff, CF24~3AA, UK}

\altaffiltext{3}{Jet Propulsion Laboratory, Pasadena, CA 91109-8099}

\altaffiltext{4}{Department of Physics \& Astronomy, University of British Columbia, 6224 Agricultural Road, Vancouver,
 BC V6T~1Z1, Canada}

\altaffiltext{5}{Department of Physics and Astronomy, University of Pennsylvania, 209 South 33rd Street, Philadelphia, PA 19104}

\altaffiltext{6}{Department of Physics, University of Miami, 1320 Campo Sano Drive, Coral Gables, FL 33146}

\altaffiltext{7}{Instituto Nacional de Astrof\'isica \'Optica y Electr\'onica (INAOE), Aptdo. Postal 51 y 72000 Puebla, Mexico}

\altaffiltext{8}{Astrophysics Group, Imperial College London, Blackett Laboratory, Prince Consort Road, London SW7 2AZ, UK}

\altaffiltext{9}{Canadian Institute for Theoretical Astrophysics, University of Toronto, 60 St. George Street, Toronto, ON M5S~3H8, Canada}

\altaffiltext{10}{Department of Physics and Astronomy, The Open University, Walton Hall, Milton Keynes, MK7 6AA, UK}

\altaffiltext{11}{Department of Physics, University of Toronto, 60 St. George Street, Toronto, ON M5S~1A7, Canada}

\altaffiltext{12}{University of Puerto Rico, Rio Piedras Campus, Physics Dept., Box 23343, UPR station, San Juan, Puerto Rico}

\altaffiltext{13}{INAF - Osservatorio Astrofisico di Arcetri, Largo E. Fermi 5, I-50125, Firenze, Italy}

\altaffiltext{14}{Universit\'{e} Paris Diderot, Laboratoire APC, 10, rue Alice Domon et L{\'e}onie Duquet 75205 Paris, France}

\altaffiltext{15}{Department of Physics, Brown University, 182 Hope Street, Providence, RI 02912}

\altaffiltext{\dag}{\url{viero@astro.utoronto.ca}}
\begin{abstract}
We detect correlations in the cosmic far-infrared background due to the 
clustering of star-forming galaxies in observations made with the 
Balloon-borne Large Aperture Submillimeter Telescope, BLAST, at 250, 
350, and 500\rmicron.  We perform jackknife and other tests to confirm the
reality of the signal.  The measured correlations are well fit by a 
power law over scales of 5--25 arcminutes, with $\Delta I/I = 15.1 \pm 1.7\%$.  
We adopt a specific model for submillimeter sources in which the contribution to clustering comes from sources in the redshift ranges ${\rm 1.3 \le z \le 2.2,~ 1.5 \le z \le 2.7,~ and~ 1.7 \le z \le 3.2}$, at 250, 350 and 500\rmicron, respectively.
With these distributions, our measurement of the power spectrum, $P(k_{\theta})$, corresponds to linear bias parameters, $b = 3.8\pm 0.6, 3.9\pm 0.6$ and $4.4 \pm 0.7$, respectively.  
We further interpret the results in terms of the halo model,
and find that at the smaller scales, the simplest halo model fails to fit our results.  
One way to improve the fit is to increase the radius
 at which dark matter halos are artificially truncated in the model, which is equivalent to
having some star-forming galaxies at $z \ge 1$
located in the outskirts of groups and clusters.  %
In the context of this model we find a minimum halo mass required to host 
a galaxy is $\mathrm{log}(M_{\mathrm{min}}/\rm M_{\odot}) = 
11.5_{-0.1}^{+0.4}$, and we derive effective biases $b_{\rm 
eff} = 2.2 \pm 0.2, 2.4 \pm 0.2 $, and $2.6 \pm 0.2 $, and effective 
masses $\mathrm{log}(M_{\mathrm{\rm eff}}/\rm M_{\odot}) 
= 12.9 \pm 0.3$, $12.8 \pm 0.2$, and $12.7 \pm 0.2$
, at 250, 350 and 500\rmicron, corresponding to spatial correlation lengths 
of $\rm r_0 = 4.9, 5.0,$ and $5.2~\pm 0.7 ~ h^{-1}~ \rm Mpc$, respectively. 
Finally, we discuss implications for clustering measurement strategies with \emph{Herschel} and \emph{Planck}.\\
\end{abstract}

\section{Introduction}
With the discovery of the cosmic far-infrared background 
\citep[CIB,][]{puget1996, fixsen1998} and subsequent studies from the 
mid-infrared to the millimeter, it has been established that the peak 
epoch of star formation lies between $1 \lsim z \lsim 3$ 
\citep{dickinson2003, hopkins2004}.  

Where star formation occurs with respect to the underlying dark matter distribution is less well understood.
In the local Universe, the sites of most active star-formation occur far from 
the densest environments, a property which reverses at $z \sim 1$ 
\citep{elbaz2007}.  This points to the environment contributing to the 
mechanisms that trigger or quench star formation.

Measurements of the clustering of star-forming galaxies on large and small
scales can be used to directly relate star formation to 
the environment. In the regime of linear growth of structure, the clustering amplitude relative to that of the underlying dark matter is described by a bias parameter, $b$, which describes how strongly star formation traces the underlying dark matter.
On smaller scales, the distribution of galaxies hosted by a dark matter halo is described by the halo occupation distribution \citep{peacock2000a}.  It contains information on the abundance of star-forming sources within 
individual halos.  Targeting the submillimeter band is particularly 
efficient for observing star formation directly.  The submillimeter 
background results from the thermal emission of interstellar dust in 
high-redshift star-forming galaxies, which is heated by optical and 
ultraviolet radiation from stars and to a lesser extent active galactic 
nuclei \citep[see][]{blain2002}.  A substantial effort has been devoted to surveys of these galaxies 
\citep[e.g.,][]{smail1997, hughes1998, barger1998, borys2003}; thus, in principle, a clustering signal could be measured from these sources directly.  

However, direct measurement of the clustering properties of resolved 
submillimeter galaxies has been elusive.  Due to limited mapping 
speeds, the areal coverage of even the most ambitious submillimeter 
surveys has been relatively small \citep[e.g., SHADES mapped 
approximately a quarter square degree at 850\rmicron;][]{coppin2006}.  In addition, 
due to steeply falling counts and modest resolutions of single-dish 
submillimeter telescopes, source confusion has made it difficult to 
resolve any sources other than those with very high signal-to-noise.  
These sources span a relatively wide redshift range, roughly $1 \le z \le 4$, peaking at 
$z \sim 2.4$ \citep{chapman2005}, which has the effect of washing out the angular 
clustering signal, further complicating measurements.
This difficulty was confirmed by \citet{scott2006}, who re-analyzed all 
the SCUBA fields and found tenative evidence of strong angular clustering, 
but with errors too large to adequately constrain the spatial 
correlation length.  Furthermore, relatively large beams have made it 
difficult to efficiently identify direct counterparts 
\citep{barger1999, ivison2000} in order to obtain spectroscopic or 
photometric redshifts.  \citet{blain2004} attempted to measure the 
spatial clustering properties combining 73 sources with spectroscopic 
information, but again were only able to tentatively measure the 
clustering length.  
Other SCUBA galaxy clustering measurements, some tentatively detecting
clustering, have been made by \citet{webb2003a}
and \citet{blake2006}; meanwhile \citet{almaini2003} claim to have
found evidence for strong angular clustering between X-ray and
submillimeter populations, although \citet{borys2004} were unable to
confirm the result using another, seemingly less biased, estimator.
Using a nearest-neighbor analysis, \citet{greve2004} find that most
significant MAMBO ($1.2\,$mm) sources come in pairs, separated by $\sim 
23$ arcsec.
Put together, these limitations have made it 
extremely difficult to measure the clustering signal robustly.

To study the clustering properties of submillimeter galaxies it is more powerful to consider the statistics of the 
{\it unresolved\/} CIB, which contains the full intensity, rather than working from a limited catalog.  
In other words, instead of measuring a correlation among numbers of galaxies, use the background fluctuations of the total intensity to measure correlations among brightnesses of galaxies.  \citet{devlin2009} and \citet{marsden2009} have demonstrated that the CIB is composed of emission by discreet sources.  Since submillimeter galaxies are optically thin, this signal will be proportional to the total star formation rates of those sources.  
Correlations in the CIB will
have a contribution in excess of white noise -- which arises from Poisson 
sampling of a background made up of a finite number of sources -- in 
the presence of clustering, with an amplitude that should be detectable 
with current surveys \citep{scott1999, haiman2000, knox2001, 
magliocchetti2001, perrotta2003, amblard2007, negrello2007}.  Initial 
attempts to detect correlations by \citet{peacock2000b}, for the 
Hubble Deep Field observed by SCUBA at 850\rmicron, and by 
\citet{lagache2000b} for a 0.25 deg$^2$ \emph{ISO\/} field at 170\rmicron,
were only able to measure a signal consistent with the 
Poisson contribution.  More recently, \citet{grossan2007} and 
\citet{lagache2007} reported the weak detection of a clustering component in 
160\rmicron\ data from $\sim 9$ deg$^2$ \emph{Spitzer\/} fields.

In this paper
we report the detection of correlations in the submillimeter part of 
the CIB due to the clustering of 
star-forming galaxies, in a 6 deg$^2$ field centered on the Great 
Observatories Origins Deep Survey South field 
\citep[GOODS-South;][]{giav2004}.  
These data were collected by the Balloon-borne Large Aperture 
Submillimeter Telescope \citep[BLAST;][]{devlin2009}, which is designed 
to bracket the peak of redshifted thermal emission from dust by 
observing at 250, 350, and 500\rmicron.  Operating above most of the 
atmosphere, BLAST is able to make observations in bands which are 
difficult or impossible to observe from the ground.  A detailed 
description of the instrument and calibration can be found in 
\citet{pascale2008} and \citet{truch2009}. 

This paper is organized as follows.  In Part I we describe how we make 
the measurement -- from map preparation to power spectrum calculation. 
 We address each contribution to the total power spectrum, and how they 
are removed to uncover the clustering signal.  
We show that the observed spectra are consistent across BLAST bands and correspond to a modest bias.  
In Part II we assess the 
plausibility of the detection by fitting models to the data: beginning with a simple linear bias model, followed by a more detailed halo model \citep{mo1996, cooray2002} and \emph{halo occupation distribution\/} \citep*{peacock2000a}.  
When required we adopt the concordance model, a flat $\Lambda$CDM 
cosmology with $\Omega_{\rm M}=0.274$, $\Omega_{\Lambda}=0.726$, $H_0 = 
70.5$ km s$^{-1}$ Mpc$^{-1}$, and $\sigma_8 = 0.81$ \citep{hinshaw2009}.
\section*{Part I: Measuring Correlations in the CIB}
\label{sec:background}
\section{Background Correlations and the Correlation Function: Overview}
Galaxy clustering can be expressed in a number of ways, the most common 
being the two-point correlation function, $w (\theta)$, which measures 
the number of pairs at a given distance in excess of what would be 
expected of a Poisson distribution.  
Alternatively, the 
clustering of galaxies can be expressed as a power spectrum in excess of Poisson noise, $\tilde{P}(k_{\theta})$, which can be expressed in dimensionless units as the fractional variance per 
logarithmic increment of wavenumber \citep[see][]{peacock_book}, i.e.,   
\begin{equation}
\tilde{\Delta}_{k_{\theta}}^2 \equiv 2\pi k_{\theta}^2 \tilde{P}(k_{\theta}),
\end{equation}
where $k_{\theta}$ is the angular wavenumber, which is also known as $\sigma$ in the literature, and is expressed in inverse angular scale as $k_{\theta} = 1/ \lambda$.  It is related to the multipole index, $\ell$, by $\ell = 2\pi k_{\theta}$.  
The tildes over $\Delta_{k_{\theta}}^2$ and $P(k_{\theta})$ denote that these quantities refer to galaxy {\it locations} rather than intensities.

Naturally, the correlation function and the power spectrum of galaxy 
clustering are related; they form a Hankel transform pair.  
Explicitly, for small surveys, 
\newline
\small
\begin{eqnarray}
   \tilde{\Delta}^2_{k_{\theta}} & = & (2\pi k_{\theta})^2 \int ^{\infty} _0 w (\theta) J_0 (2\pi k_{\theta} \theta) 
\theta\ d\theta , \nonumber\\
  w (\theta) & = & \int ^{\infty} _0 \tilde{\Delta}^2_{k_{\theta}} J_0 (2 \pi k_{\theta} \theta)\  dk_{\theta} / 
k_{\theta}. 
  \label{eq:FTP}
\end{eqnarray}
\normalsize
For small separation correlations of local galaxies, angular clustering is often described as a power law, $w(\theta) = 
(\theta/\theta_0)^{-\epsilon}$, were $\epsilon \simeq 0.8$ is the canonical 
slope  
\citep[e.g.,][]{giav1998}.  The power spectrum analog for small areas 
\citep[see][]{peacock2000b} is 
\begin{equation}
\tilde{\Delta}^2_{k _{\theta}} (k_{\theta}) = (2\pi k_{\theta} \theta_0)^{\epsilon} 2^{1-\epsilon} 
\frac{\Gamma(1- \epsilon/2)}{\Gamma(\epsilon/2)}, 
\label{eq:Delta2}
\end{equation}
which is equal to $3.35(k_{\theta}\theta_0)^{0.8}$ for $\epsilon \simeq 0.8$.

As previously stated, this holds for correlations of galaxy \emph{locations}.  On the other hand, the power spectrum of background fluctuations, which is what we calculate, measures correlations of galaxy \emph{intensities}. 
The redshift distribution of the cumulative flux contributed by the background sources is represented as 
\begin{equation}
  \frac{d{\cal S}}{dz} = \int_{0}^{\infty}S\frac{dN}{dS\ dz}(S,z)\ dS,
\end{equation}
where $dN/(dSdz)$ is the number density of sources per unit flux
density and redshift interval.
The measured power spectrum of background fluctuations, $P(k_{\theta})$, is 
the 2-dimensional, \emph{flux weighted}, projection of the 
3-dimensional galaxy clustering spectrum, $P_{\rm 3D}(k_{\theta})$.  For $k_{\theta}\gg1$ and a flat cosmology, 
their relationship can be approximated as
\begin{equation}
\small
  P(k_{\theta}) = \int_{z_{\rm min}}^{z_{\rm 
max}}P_{\rm 3D}(2\pi k_{\theta}/x(z),z)\left(\frac{d{\cal 
S}}{dz}(z)\right)^{2}\frac{1}{dV_{\mathrm{c}}(z)}\ dz,
  \label{eq:Cell_vs_Pk}
\normalsize
\end{equation}
\citep[e.g.,][]{tegmark2002}, where $x(z)$ is the comoving radial 
distance and $dV_{\mathrm{c}}(z)$ is the comoving volume element, i.e., 
$dV_{\mathrm{c}}(z) = x(z)^2dx/dz$ (assuming a flat cosmology), 
where $dN/(dSdz)$ is the number density of sources per unit flux 
density and redshift interval.

$P(k_{\theta})$ can expressed in dimensionless units as
\begin{equation}
\Delta_{k_{\theta}}^2 \equiv 2\pi k_{\theta}^2 P(k_{\theta})/I^2_{\nu},
\label{eq:Delta2_2}
\end{equation}
where $I_{\nu}$ is the intensity of the background in Jy.  Although $\Delta_{k_{\theta}}^2$ and $\tilde{\Delta}_{k_{\theta}}^2$ have the same units and similar meaning, they differ because Equation~\ref{eq:Delta2_2} deals with an intensity weighted power spectrum.

In addition to the clustering signal, the total power spectrum has 
contributions from instrumental noise, Poisson noise from individual 
background galaxies, and cirrus emission.  We will address each 
contribution individually. 

\section{Methods}
\label{sec:methods}

\subsection{Map Preparation}
\label{sec:map}
BLAST observed a wide $ 8.7 $ deg$^2$ patch, centered on the 
GOODS-South field ($3^{\rm h} 32^{\rm m} 35^{\rm s}, -28^{\circ} 15\arcmin$; hereafter BGS-Wide), with mean 1-$\sigma$ sensitivities of
36, 31, and $20\,{\rm mJy\,{\rm beam}^{-1}}$  at 250, 350, and 500\rmicron,
as well as a deep, nested field of $0.8$ deg$^2$, centered on
($3^{\rm h} 32^{\rm m} 30^{\rm s}, -27^{\circ} 48\arcmin$; hereafter BGS-Deep) with mean
1-$\sigma$ sensitivities of 11, 9 and $6\,{\rm mJy}\,{\rm beam}^{-1}$,
respectively\footnote[1]{BLAST maps and catalogs are 
publicly available at http://www.blastexperiment.info}.  A $6 $ deg$^2$ 
region, centered on BGS-Wide, is selected from the map, because of its 
uniformity in observed depth.  The BLAST bolometers are prone to drifts on 
timescales greater than $\sim 10$ seconds.  %
To retain as much large angular scale signal as possible, a fast scan rate is preferred 
because larger scales will survive the high-pass filtering, designed to remove noise below the $0.1\ \rm Hz$ $1/f$ knee.  For this 
analysis we use only the data from the wide region of the map, where 
the scan rate is $0.1\,{\rm deg}\,{\rm s}^{-1}$, and the r.m.s. of the maps
is 2.4, 1.9, $1.0\,{\rm MJy}\,{\rm sr}^{-1}$, at 250, 350, and 500\rmicron, which is applicable to calculating uncertainties of \emph{point-source} flux densities.
The data for the nested deep region, whose scan rate is only
$0.05\,{\rm deg}\,{\rm s}^{-1}$, are not included. 
Large-scale noise is removed by high-pass filtering the time-streams at 
0.2 Hz.  Correlated noise -- a drift of multiple detectors in unison 
-- is not removed because it would inevitably suppress large scale 
signal as well.  Instead, a cross-correlation of a subset of the maps 
is used to remove large-scale correlated noise (described below).  
Although fully optimized map-makers are available \citep[e.g., 
SANEPIC;][]{patanchon2008}, the long time to convergence (typically 24 
hours on 10 processors for this particular map) makes it impractical to 
use them for our Monte-Carlo simulations.  Full analysis with SANEPIC 
maps, including the nested deep field, will be the subject of a future 
paper.  Here we instead use OPTBIN \citep{pascale2009} a fast, naive, 
map-maker whose transfer function is calculated using a Monte-Carlo 
simulation  (see \S~\ref{sec:power_spectrum} and Figure~\ref{fig:tf} 
for details), and we use SANEPIC maps for consistency checks.  Parallel 
power spectrum analysis with both map-makers show agreement well within 
the simulated uncertainties. 

\subsection{Power Spectrum Calculation}
\label{sec:power_spectrum}
The map intensity, $S_{\mathrm{map}}$, can be written as
\begin{equation}
S_{\mathrm{map}} = \big{(}T\otimes[S_{\mathrm{sky}}\otimes B + 
N]\big{)}\ W,
\end{equation}
where we use $\otimes$ to represent a convolution, $S_{\mathrm{sky}}$ is the true sky surface brightness, $T$
is the transfer function of the map-maker, $B$ is the measured 
instrumental beam, $N$ is the instrumental noise, and $W$ is 
the \lq aperture function\rq, which is zero beyond the region of interest.  
The autocorrelation of a map will contain a contribution from detector noise.  To suppress this instrumental noise, 
cross-power spectra are taken among a set of four maps which are made by 
dividing the time-stream into four roughly equal parts and then making
four separate maps (hereafter referred to as sub-maps).  The 
timestreams, which are made up of numerous chunks, are divided into 
every fourth chunk (e.g., 1, 5, 9, \dots, and 2, 6, 10, \dots, etc.), so that 
the sub-maps have as similar coverage as possible.  The number of 
sub-maps chosen maximizes the number that can be made while maintaining 
uniformity in hits, retaining some cross-linking, and avoiding holes in 
the maps.  The r.m.s. of the resulting sub-maps is
4.6, 3.6, and $2.0\,{\rm MJy}\,{\rm sr}^{-1}$, at 250, 350, and 500\rmicron.
In the cross-spectrum, 
noise which is uncorrelated between sub-maps averages to zero.  
Consequently, the spectrum does not depend on modeling the potentially 
complicated or non-stationary noise.  

\begin{figure}
\centering
\vspace{0.1cm}\includegraphics[angle=270,width=.90\linewidth]
{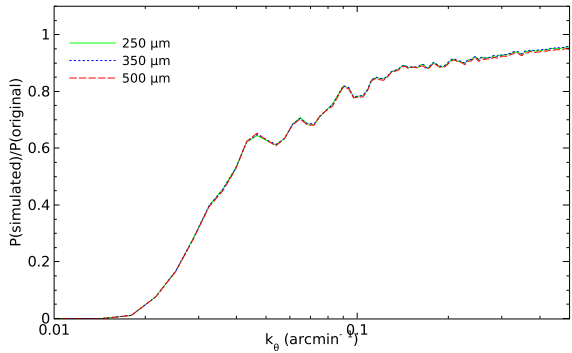}
\caption{Transfer functions calculated with a Monte-Carlo simulation 
involving 500 mock-maps observed with the BLAST simulator.  The lines 
are identical beyond the scale of the beams, which implies that the 
map-maker is linear, as we would expect.}
\label{fig:tf}
\end{figure}

We prepare the maps before calculating the power spectrum by removing 
their means, apodizing them with a Welch window \citep[][chapter 
13.4]{nr2002}, and zero-padding them with a width on each side equal to 
half the map.  The cross-correlation two-dimensional power spectrum of 
each pair of maps is calculated.  The azimuthal average of the 
amplitudes (which in two-dimensional $k$-space appears to be isotropic),
is taken to find the one-dimensional power spectrum, $P(k)$.  
The resulting spectra are averaged and divided by the power spectrum of 
the beam and the transfer function.  The transfer function is 
calculated with a Monte-Carlo simulation from simulated maps made with 
the BLAST simulator\footnote[2]{We simulate the BLAST pipeline by \lq 
observing\rq\ an input map, creating a set of time-streams, filtering them,
and making a map from the filtered timestreams.}, and is shown in 
Figure~\ref{fig:tf}.
At angular scales which are large compared to the high-pass filter ($< 
0.03$ arcmin$^{-1}$) and approaching the cut-off of the beam ($> 
0.9$ arcmin$^{-1}$), the transfer function is unreliable.  We are 
interested in the scales bracketed by these limits.   

\subsubsection{Jackknife tests}
\label{sec:jacknife}
We have performed jackknife tests in which a correlation is calculated between two distinct difference maps.  
Specifically, we take the cross-correlation of: 
$(\mathrm{sub}$-$\mathrm{map}\ 1 - \mathrm{sub}$-$\mathrm{map}\ 2)$ 
and $(\mathrm{sub}$-$\mathrm{map}\ 3 - \mathrm{sub}$-$\mathrm{map}\ 
4)$.  If the cross-correlation measures only signal, then taking a difference between sub-maps should cancel sky signal and result in a cross-correlation power spectrum consistent with zero.  And indeed, our results are consistent with zero, in the range of interest, 
in all three bands.

As a further check, the cross-power spectrum is compared to the 
difference of the auto-power spectrum (of a map made from the entire 
timestream) and the measured noise.  The noise is estimated from 
odd-even pixel jackknife maps, and is approximately white over the 
angular scales of interest.  The resulting two spectra are in excellent 
agreement.  

\newpage
\subsubsection{Poisson Noise}
Poisson (or shot) noise arises from the finite number of galaxies per 
unit area.  It can be calculated analytically from the source counts 
as
\beq
P_{\mathrm{shot}} = \int_0 ^{\infty} S^2 \frac{dN}{dS} dS.
\eeq
Alternatively, Poisson noise levels can be estimated from Monte-Carlo 
simulations using mock-maps which are populated with uncorrelated 
sources whose fluxes are drawn from the measured BLAST counts 
\citep{patanchon2009}.  This has the added advantage that realistic 
uncertainties can be calculated as well.  Since the counts are so 
steep, care must be taken to reproduce the bright end of the counts 
faithfully, so that extremely rare bright sources do not 
unrealistically appear in the realizations.  We find that the two 
methods of estimating the level of shot noise
agree within the 1-$\sigma$ uncertainties.

\begin{figure}
\centering
\includegraphics[width=1.\linewidth]
{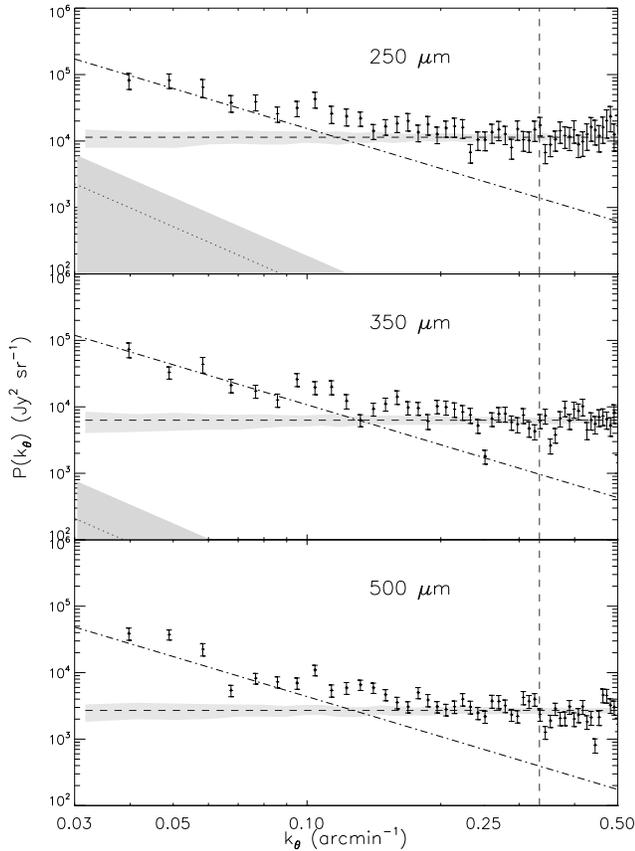}
\caption{Power spectra of 6 deg$^2$ regions selected from the BGS-Wide 
maps at 250, 350, and 500\rmicron\ are shown with 1-$\sigma$ 
uncertainties.  Color-corrected fits to galactic cirrus measured from
 IRIS 100\rmicron\ maps are shown as dotted lines with shaded region 
representing 1-$\sigma$ uncertainties, sloping down from the left 
side  (500\rmicron\ cirrus is too low to make it onto the region
plotted).  Poisson noise contributions 
are found with Monte-Carlo simulations, and are shown as horizontal 
dashed lines with similar error regions. Scale invariant, $k^{-2}$, power spectra are shown as dot-dashed lines.  Vertical dashed line at $k_{\theta} = 0.33$ arcmin$^{-1}$ represents the scale at which the variance in the FIDEL catalog --- which \citet{marsden2009} show resolves most of the CIB --- is no longer Poisson noise dominated.  
For angular scales greater than 0.33 arcmin$^{-1}$, signal in excess of Poisson noise is attributed to the clustering of star-forming galaxies.
}
\label{fig:plot_all}
\end{figure}

Due to the steep nature of the source counts, the Poisson noise is 
dominated by the contributions of the fainter population.  Furthermore, \citet{marsden2009} find only 15\% of the total sky intensity is associated with a 3-$\sigma$ catalog.  We find that 
removal of only the brightest sources results in an approximately 5\% 
reduction in Poisson noise at 250\rmicron, and that more aggressive 
cuts lead to removal of correlations beyond just Poisson noise.  We 
find behavior consistent with this in our simulations.  Therefore, we subtract 
only 5 sources above $500\,$mJy at 250\rmicron, 2 sources above $400\,$mJy 
at 350\rmicron, and no sources at 500\rmicron.  To subtract sources, 
first we make a source list by performing a noise-weighted convolution 
of the maps with the effective BLAST point-spread function (PSF) and 
identify local maxima in the smoothed map.  We then subtract a scaled 
effective PSF with amplitude taken from the source list.  We perform the 
same operation on our mock-maps, from which we calculate Poisson levels 
of $11.4 \pm 1.0 \times 10^3, 6.3 \pm 0.5 \times 10^3$, and $2.7 \pm 
0.2 \times 10^3{\rm Jy}^2{\rm sr}^{-1}$ at 250, 350, and 500\rmicron.  These 
are shown as dashed lines in Figure~\ref{fig:plot_all}.

\subsubsection{Estimates of Galactic Foregrounds}
\citet{gautier1992} show that the power spectrum of Galactic cirrus can be 
approximated by a power law,
\begin{equation}
P_{\mathrm{cirrus}}(k_{\theta}) = P_0\left( \frac{k_{\theta}}{k_0} 
\right) ^{-\alpha},
\end{equation}
where $k_{\theta}$ is the angular wavenumber in inverse arcminutes, and 
$P_0$ is the power spectrum value at $k_0 = 0.01$ arcmin$^{-1}$.  We 
measure the cirrus component at 100\rmicron\ from the 
cross-correlation of the three co-added IRIS (HCON\footnote[3]{HCON 
refers to each survey.  For more information and maps see 
http://www.cita.utoronto.ca/~mamd/IRIS/IrisOverview.html}) maps 
\citep[reprocessed \emph{IRAS\/}:][]{miv2005} for a $\sim 15$ deg$^2$ 
region surrounding the BGS-Wide field.  
The amplitude of the observed power spectrum has a contribution from cirrus emission which is highly variable on the sky.  The GOODS region was specifically chosen because it is a low cirrus region, with a mean 
intensity of $1.39\pm 0.18\,{\rm MJy}\,{\rm sr^{-1}}$.
At scales $0.008 < k_{\theta} < 0.03\,{\rm arcmin}^{-1}$, which is larger than the scales probed by BLAST, we find that the cirrus is well approximated by a power law,
$P_0 = (0.47 \pm 0.18) \times 10^6\,{\rm Jy}^2{\rm sr}^{-1}$
and $\alpha = 2.91\pm 0.11$, values which are similar to those found in low cirrus
emission regions 
measured by \citet{lagache2007} and \citet{miv2007}.  We assume the power 
spectrum continues to smaller angular scales, and scale it to the BLAST 
bands using the average dust emission color 
$(I_{\mathrm{BLAST}}/I_{100})^2$, which is found by assuming cirrus 
emission behaves as a modified blackbody, $\nu^{\beta}B(\nu)$, where 
$B(\nu)$ is the Planck function, and $\beta$ is the emissivity index 
\citep{draine1984}.  The scaled power spectrum and errors are 
calculated with a Monte-Carlo simulation, varying temperature ($17.5 
\pm 1.5$ K) and $\beta$ ($1.9 \pm 0.2$) \citep{boulanger1996}.  The 
resulting power law approximations and uncertainties are illustrated as 
dotted lines in Figure~\ref{fig:plot_all}, which show that we are 
negligibly affected by Galactic cirrus on all recovered angular scales. 

\newpage
\subsubsection{Uncertainties}
To estimate uncertainties in our angular power spectra, clustered 
signal plus noise simulated maps are analyzed with the same 
pipeline as the astronomical data.  We include clustering in the simulations in case the amplitude or shape of the transfer function depends on the input, but we found that not to be the case.  We follow \citet[][see appendix for algorithm]{almaini2005} to introduce correlations to the simulated maps, 
with an angular correlation length, $\theta _0 = 3.0\arcsec$.  This angle was chosen 
from the measured upper limit of \citet{peacock2000b}.  Realizations with stronger clustering clearly do not \emph{look} like BLAST maps.  
\section{Basic Clustering Results}
\label{sec:results1}
\begin{table*}[ht]
  \centering
  \begin{tabular}{lccc}
   \hline
   \hline
$k_{\theta}$ & BLAST 250 & BLAST 350 & BLAST 500 \\
$(\rm arcmin^{-1})$ & $(\rm steradian^{-1})$ & $(\rm steradian^{-1})$& $(\rm steradian^{-1})$\\
   \hline
     0.044 &  $(1.39\pm 0.52 ) \times 10^{-7}$ & $(1.33\pm 0.47 ) \times 10^{-7} $ & $(2.47\pm  1.17 ) \times 10^{-7}$\\
     0.070 &  $(5.94\pm 2.02 ) \times 10^{-8}$ & $(4.98\pm 1.83 ) \times 10^{-8} $ & $(5.71\pm  2.06 ) \times 10^{-8}$\\
      0.113 & $(2.99\pm 0.95 ) \times 10^{-8}$ & $(2.62\pm 0.90 ) \times 10^{-8} $ & $(2.98\pm  1.00 ) \times 10^{-8}$\\
      0.183 & $(9.81\pm 7.51 ) \times 10^{-9}$ & $(9.94\pm 6.92 ) \times 10^{-9} $ & $(6.91\pm  4.62 ) \times 10^{-9}$\\
   \hline
  \end{tabular}
  \caption{CIB normalized clustering power spectra, $P(k_{\theta})/I^2_{\nu}$.  The errors do not include uncertainties in the CIB.  CIB values are listed in \S~\ref{sec:results1}}
  \label{tab:results}
\end{table*}
Figure~\ref{fig:plot_all} shows an unambiguous signal in excess of
Poisson noise on scales of $0.04 - 0.2$ arcmin$^{-1}$ which cannot be
explained by Galactic cirrus, and which we interpret as correlations
from clustered star-forming galaxies.  The vertical dashed line indicates the angle at which the distribution of 24\rmicron\ selected FIDEL galaxies \citep[][]{magnelli2009} begin to show variance in excess of Poisson.  Indeed, the similarity to Figure 3 of \citet{marsden2009} is striking.

\begin{figure}
\centering
\includegraphics[width=1.0\linewidth]
{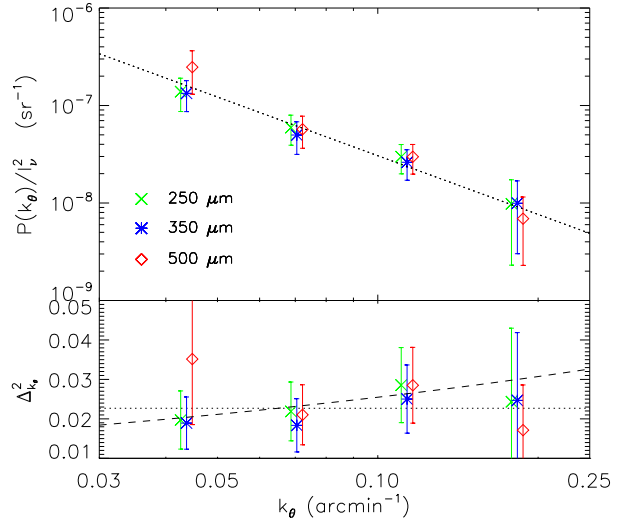}
\caption{Top: 
Power Spectra of the clustering component, after removal of cirrus and Poisson noise, and normalized by ($I_{\nu}^{\rm CIB})^2$.
The best-fit 
power spectrum,  proportional to $k^{-2}$, is shown as a dotted line.
Bottom: $\Delta^2_{k_{\theta}}$ shown along with best-fit 
power spectra (horizontal dotted line); as well as best-fit parameters to Equation~\ref{eq:Delta2} (dashed line) for 250\rmicron\ data.  The 250 and 500\rmicron\ points are offset horizontally for clarity, by factors of -0.025 and +0.025, respectively.  %
Clearly, the power spectrum signal of clustered star-forming galaxies is well fit by a power law 
power spectrum proportional to $k^{-2}$.  
}
\label{fig:PSwide}
\end{figure}

The CIB has an amplitude $I_{\nu}^{\rm CIB}$ measured to be
$0.71\pm0.17$, $0.59\pm0.14$, and $0.38\pm0.10\,{\rm MJy}\,{\rm
  sr}^{-1}$ at 250, 350, and 500\rmicron, respectively
\citep[][]{marsden2009}.  Figure~\ref{fig:PSwide} shows the
clustering component of the power spectrum normalized by $I_{\nu}^{\rm
  CIB}$ in the three BLAST bands, where the 250\rmicron\ and 500\rmicron\ data have
been displaced slightly for visual clarity.  The contributions of cirrus and
Poisson noise have been subtracted, and the data rebinned in
logarithmic intervals.  These results are also listed in
Table~\ref{tab:results}.  We use the BLAST estimates for the CIB
because they are the most precise estimate available of the CIB in
these wavelength bands.  Doing so has the additional benefit in this
case that calibration uncertainties completely vanish in the ratio.
The fit to a single power law, and also the agreement in amplitude
across the BLAST bands, once the power spectra are normalized to the
sky intensity, are excellent.

The relative variance of the CIB, $\Delta^2_{k_{\theta}}$, 
formed by dividing $P(k_{\theta})$ by $2\pi k_{\theta}^2$, is shown in the bottom panel
of Figure~\ref{fig:PSwide}, where the 250\rmicron\ and 500\rmicron\ data have
been displaced slightly for visual clarity. (Amplitudes have been
additionally converted from $\rm arcmin^{-2}$ to $\rm steradian^{-1}$ in the
figure and in the discussion below.)  %
The best fits of Equation~\ref{eq:Delta2}) are shown in Table~\ref{tab:fit_delta}. 
The data are consistent (to within 1$-\sigma$) with $\Delta^2_k={\rm Constant}$ 
, with the same amplitude in all three bands
, even taking into account that the error bars shown contain a
large component of uncertainty which is common mode between channels,
and thus overestimate the anticipated scatter.  The best-fit 
amplitude for a power law with slope of $-2$ is shown as a dotted line in the bottom panel, corresponding to
 $\Delta_{k_{\theta}} = \delta I/I= 15.1 \pm 1.7\% $.  This is directly
 analogous to the square root of the
`band-power' measured in Cosmic Microwave Background (CMB) anisotropy
experiments, where typically $\delta T/T\sim10^{-5}$.  From this point of view
the CIB is much clumpier than the CMB, as one would expect since the
galaxies are observed after a much longer period of linear growth.
The dotted line in the upper panel of Figure~\ref{fig:PSwide} is
$2\pi k_{\theta}^2\Delta^2_{k_{\theta}}$, corresponding exactly to the fit in the
lower panel.  

The values of the parameters $\epsilon$ and $\theta_0$ (see Equation~\ref{eq:Delta2}) which best fit the data are shown in Table~\ref{tab:fit_delta}.  The large uncertainty quoted for $\theta_0$ is due partly to the awkwardness of this parameterization near to zero slope.  The bottom panel of Figure~\ref{fig:PSwide} shows the best-fit for 250\rmicron\ data as a dashed line.  
Such a small $\theta_0$, despite significant power in excess of Poisson noise, requires that the sources which make up the CIB are distributed over a wide range of redshifts, and has implications for future clustering measurement strategies (e.g., with \emph{Herschel} or \emph{Planck}), which we discuss in \S~\ref{sec:conclusions}.  

We have calculated the correlations between different bands (e.g., $250 
\times 350$) using the same pipeline and set of sub-maps.
 The cross-spectra are  normalized by the square root of the
auto-spectra of the two bands, so that the final curve would be unity
at all scales for identical maps, and zero at all scales for two
completely different maps.  Results show that the
cross-correlations are  $0.95\pm 0.06$, $1.06\pm 0.09$, and $0.92\pm 0.04$, for $250 
\times 350$, $350 \times 500$, and $250 \times 500$, respectively, over
the range of angular scales $0.04 < k_{\theta} < 0.5$ arcmin$^{-1}$.
Neighboring bands are more correlated with each other than are the
250 and 500\rmicron\ bands.  While we find the same
spectrum in all three bands, the phases are different, as one would
expect if the three BLAST bands have different selection functions and
sample the galaxies in the CIB at different redshifts.  While the cross-band
correlation provides a powerful tool for testing the redshift
distributions and spectral energy densities of source population
models, a more detailed analysis is required before making any strong
conclusions.  This will be the subject of a future study using the
SANEPIC maps and including the BGS-Deep data.


We have measured the variance projected onto two dimensions, but
galaxies are of course distributed in three dimensions.  Knowledge of
the redshift distribution of the galaxies in the CIB allows
interpretation of these power spectra in terms of a bias factor
quantifying the comparison to cold dark matter (CDM) spectra, which we find in the next
section.  As the angular scales we probe get smaller, non-linear growth eventually sets in, 
which we examine by fitting halo models to these spectra later in the paper.
%
%
\begin{table}
  \centering
  \begin{tabular}{lcc}
   \hline
   \hline
   BAND         &  $\theta_0$ (arcsec) & $\epsilon$       \\
   \hline
   250\rmicron  &  $0.017\pm0.020$     & $0.27\pm0.19$  \\
   350\rmicron  &  $0.008\pm0.006$     & $0.26\pm0.19$  \\
   500\rmicron  &  0.0                 & 0.0            \\
   \hline
   \end{tabular}
   \caption{Best-fit values obtained for $\theta_0$ and $\epsilon$.
}
   \label{tab:fit_delta}
\end{table}

\newpage
\section*{Part II: Model Fitting}
\section{Introduction}
\label{sec:part2}
To interpret our detection of correlations requires
comparison to an underlying model whose parameters the data constrain.
Such a model could contain details of the complete source population,
including number counts, {\it i.e.} intensity distributions, and
redshift distributions, as well as a framework for describing the
linear and non-linear clustering regimes.  This latter part might
involve, for example, a halo occupation distribution which accounts
for the galaxy distribution in a given dark matter halo as a function
of luminosity \citep[i.e., the so-called conditional luminosity
function;][]{cooray2006}.  For a background of primarily unresolved
sources, the conditional luminosity function model is an improvement
on the simple halo occupation distribution; however, its increase in
complexity comes with an increase of free parameters.

Given the very good fit to a constant $\Delta^2_{k_{\theta}}$, we first explore the
physical meaning of the BLAST power spectra in the context of a purely
linear model.  We assume that
 the galaxies which comprise the CIB  have a power spectrum which is
 scaled from the power spectrum of dark matter at every redshift,
$P(k)=b^2 P_{\rm DM}(k)$.  Because  $P_{\rm DM}$ is redshift
dependent, estimating $b^2$ requires knowledge of the redshift
distribution of the galaxies which comprise the BLAST signals.
 To find this distribution  we
adopt the model of \citet{lagache2004}, described in
Section~\ref{sec:source_pop}.   It is worth noting that since we are
really measuring an emission-weighted bias at rest-frame far-IR
wavelengths, there will be some degeneracy between the value of $b$
and changes in the redshift distribution or far-IR spectral shapes
assumed for the sources.

We extend the fit into the non-linear regime to
study whether the BLAST data can constrain  parameters of the
halo occupation distribution.  We can check whether our correlations
are consistent with the model (e.g., by judging whether the same bias
fits all three BLAST wavebands, and that the cross-band correlations of
simulations made with the model agree with the data), but we will not
explore how the model might be improved.
It is important to understand that if the distribution of source
redshifts or counts were different, then we would infer a different
bias level.  However, since the model we've adopted agrees with a large body of
multi-wavelength observations, we are confident that these prescriptions
are a reasonable first attempt.

\section{Source Population Model}
\label{sec:source_pop}
\begin{figure}[!t]
  \hspace{0.0cm}\vspace{0.0cm}\includegraphics[height=5.7cm,width=9cm]
  {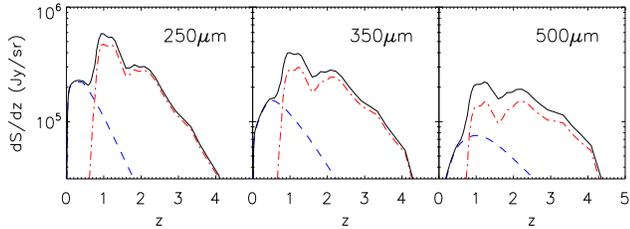}
  \vskip-2.5truecm\caption{Redshift distribution of the cumulative flux 
contributed by the background sources at the BLAST bands, according to 
the \citet{lagache2004} model. The dashed (blue) and dot-dashed (red) 
curves are for \lq regular\rq\ and \lq star-forming\rq\ \emph{IRAS\/} 
galaxies respectively, while the solid line is the total. 
$~~~~~~~~~~~~~~~~~~~~~~~~~~~~~~~~~~~~~~~~~~~~~~~~~~~$}
  \label{fig:dSdz_LagacheModel}
\end{figure}
\begin{figure}
  \hspace{0.0cm}\vspace{0.0cm}\includegraphics[height=5.9cm,width=9cm]
   {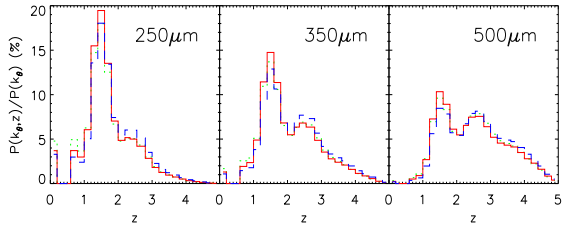}
  \vskip-2.5truecm\caption{Contribution to the angular power spectrum from different redshift slices, inferred from Figure~\ref{fig:dSdz_LagacheModel}, at specific angular scales probed by the data.
The different curves correspond to the following scales: dotted (green) -- $\rm k_{\theta} = 0.04\ arcmin^{-1}$; solid (red) --  $\rm k_{\theta} = 0.1\ arcmin^{-1}$; dashed (blue) --  $\rm k_{\theta} = 0.18\ arcmin^{-1}$.  Whereas the redshift distribution of sources (Figure~\ref{fig:dSdz_LagacheModel}) has a significant contribution from $z < 1$, the contribution from those sources to the angular power spectrum is negligible.
}
  \label{fig:rds}
\end{figure}
\begin{figure*}[t!]
\hspace{0.0cm}\vspace{0.0cm}\includegraphics[height=12.0cm,width=18.5cm]
  {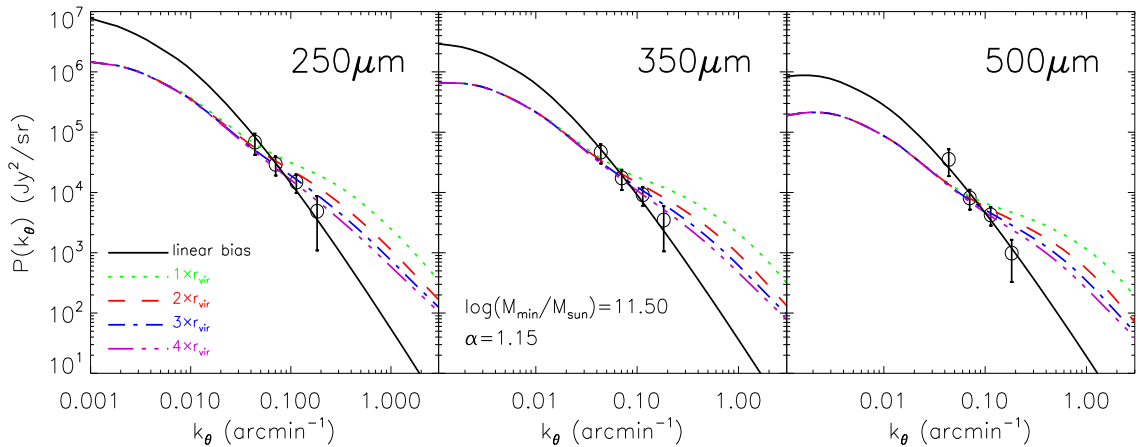}
  \vskip-4.7truecm\caption{Power spectrum of background correlations 
(circles with error bars) overlaid with the best-fit linear bias (solid line), as well as predictions obtained for halo models different values of $r_{\rm cut}$.  See Table~\ref{tab:chi2} for the fit parameters.  %
Although the power spectra shown here have different amplitudes in the three channels, $P(k_{\theta})/I_{\nu}$ is the  same in all three bands, as shown in Figure~\ref{fig:PSwide}.  The data are fit best by a model which includes a linear term only.}
  \label{fig:Pkth_data_vs_model}
\end{figure*}
Knowledge of the full redshift distribution of BLAST sources would allow us to estimate the redshift 
distribution of the measured clustering signal, i.e., the redshift range probed by the power spectrum of correlations due to clustering described by the $P\times dS/dz$ distribution.
\citet{devlin2009}, and in more detail \citet{patanchon2009}, find the
number counts of the BLAST sources, and \citet{pascale2009} divides
the redshift distribution roughly into four redshift bins.  Precisely understanding the redshift
distribution of those sources is a work in progress (e.g., Chapin et
al., in prep.).  In the meantime we adopt the model of
\citet{lagache2004} -- a model which approximates the counts and
redshift distributions of the populations expected to make up the CIB
-- to describe the underlying source
population\footnote[4]{Documentation and {\it IDL} files can be found at
  http://www.ias.u-psud.fr/irgalaxies/model.php}.  
Specifically, it is a
phenomenological model which extrapolates the local 60\rmicron\
luminosity function of \emph{IRAS\/} sources (divided into \lq
regular\rq\ and \lq star-forming\rq\ components) to longer
wavelengths, assuming a set of spectral energy distribution templates.
The extrapolation to higher redshift is parameterized as a mixture of
luminosity and density evolution, and is constrained to be consistent
with most of the available data on source counts, redshift
distributions and the far-infrared background intensity.  It becomes
less reliable with greater extrapolation to longer wavelengths, and
therefore should be considered an approximation; however, it does reproduce the BLAST \citep{patanchon2009} counts at a level which is sufficient for our purposes.  Furthermore, since the clustering signal is dominated by the faint source population, it is critical
that the model is consistent with the level of the background at the
BLAST wavelengths, and that the cross-band correlations of simulated
maps agree with those measured, which we find to be the case.

Most of the clustering signal is coming from relatively
high redshifts. The medians of the redshift distributions 
in Figure~\ref{fig:rds} are  z = 1.61, 1.88 and 2.42, at 250,
350 and 500\rmicron, respectively, and the upper
and lower quartiles of the distributions are $z=$(1.3, 2.2),
(1.5, 2.7), (1.7, 3.2), respectively.  At the representative scale of 
0.1 arcmin$^{-1}$ (red solid line in Figure~\ref{fig:rds}), 95\% of the background originates from
sources $1.1 < z$, $1.2 < z$, and $1.4 < z$ in the three bands.
\section{Linear Bias Model}
\label{sec:linear}
We first carry out a simple fit to a scaled version of the linear theory power spectrum
of the dark matter $P_{\rm DM}$.  
As can be seen in Figure~\ref{fig:Pkth_data_vs_model}, a simple
biasing prescription provides a good fit to the BLAST data.  The
required bias levels are $b=3.8 \pm 0.6$, $3.9 \pm 0.6$, and $4.4 \pm 0.7$ at 250, 350
and 500\rmicron, respectively, with a reduced $\chi^{2}_{\rm min} \sim 0.4$ (with 10 degrees of freedom) in all three bands. 

More detailed modeling could be attempted.  In principle the
cross-band measurements and wavelength dependence of the measured
correlation amplitudes could be used to estimate the variation of bias
with redshift, as discussed by \citet{knox2001}.  However, we leave
this to a future study.

\section{Halo Model}
\label{sec:halo}
As illustrated in Figure~\ref{fig:Pkth_data_vs_model}, a simple biasing
prescription provides a good fit to our data.  
The main drawback of this simple fit is that 
it may not realistically account for the 1-halo, nonlinear clustering component.  Our data are expected to bracket the physical scales corresponding to the transition from linear to non-linear clustering regimes, and though these two contributions appear to have combined to look very much like a scaled linear bias, to conclude that there is no contribution from a 1-halo term may be unphysical.  
Therefore, in this section we use a particular implementation of the \lq halo model\rq, which assigns galaxies to halos as a function of mass, and consequently probes into the territory of non-linear fluctuations, to explore how the 1-halo term could contribute power on small scales.  This also allows us to discuss our results in the context of other measurements of galaxy clustering.

The halo model of large scale structure has proven to be a powerful 
tool for describing the clustering properties of cosmic objects 
\citep[for a review, see][]{cooray2002}.  Its main ingredient is the 
parameterization of the {\it halo occupation distribution\/}
\citep[HOD,][]{peacock2000a}, which describes how galaxies populate 
dark matter halos as a function of halo mass. The power spectrum of 
galaxies is written as the sum of two components: the 1-halo term, 
$P_{\rm 1h}$, which describes pairs of objects within the same dark 
matter halo, and the 2-halo term, $P_{\rm 2h}$, which accounts for 
pairs of objects in different halos, resulting in $P(k,z) = P_{\rm 
1h}(k,z) + P_{\rm 2h}(k,z)$.  The number of pairs of galaxies within an 
individual halo is related to the variance of the halo occupation 
distribution, $\sigma^{2}( M,z) = \langle N_{\rm gal}(N_{\rm gal}-1) 
\rangle $, while the number of pairs of galaxies in separate halos is 
simply the square of the mean halo occupation number, $N(M,z) = \langle 
N_{\rm gal} \rangle$ (HON, hereafter). We model the HON using a 
central-satellite formalism \citep[see e.g.][]{zheng2005}: this assumes 
that the first galaxy to be hosted by a halo lies at its center, while 
any remaining galaxies are classified as \emph{satellites\/} and are 
distributed in proportion to the halo mass profile. Different HODs for 
central and satellite galaxies are then applied.
For central galaxies, the mean HON, $N_{\rm cen}$, is described by a 
step function such that halos above a minimum mass threshold $M_{\rm 
min}$ contain a single central galaxy and halos below this threshold
contain no galaxies. For satellite galaxies, a power-law in mass 
describes %
their mean HON \citep[e.g.,][]{zehavi2005}
\begin{table}
  \centering
  \begin{tabular}{c|ccc}
   \hline
   \hline
   $r_{\rm cut}/r_{\rm vir}$  & $\chi^{2}_{\rm min}$ &  
$\log(M_{\rm min}/M_{\odot})$  &  $\alpha$  \\
   \hline
   1  &  16.3  & $11.50_{-0.05}^{+0.40}$  &  $0.95_{-0.95}^{+0.05}$  \\
   2  &  13.6  & $11.50_{-0.05}^{+0.40}$  &  $1.00_{-1.00}^{+0.10}$  \\
   3  &  11.5  & $11.50_{-0.05}^{+0.40}$  &  $1.10_{-1.10}^{+0.05}$  \\
   4  &  9.7  & $11.50_{-0.05}^{+0.40}$  &  $1.15_{-0.75}^{+0.05}$  \\
   \hline
  \end{tabular}
  \caption{Best-fit values obtained for $M_{\rm min}$ and $\alpha$ for 
different choices of the radius $r_{\rm cut}$. The minimum-$\chi^{2}$ 
(with 10 degrees of freedom) are also shown.} 
  \label{tab:chi2}
\end{table}
\small
\begin{equation}
  N_{\rm sat}(M) = \left( \frac{M}{M_1} \right) ^{\alpha},
\end{equation}
\normalsize
where $M_1$ is the mass-scale at which a halo hosts exactly one 
satellite galaxy (in addition to the central galaxy).  Both 
semi-analytic models \citep[e.g.,][]{berlind2000} and hydrodynamical
simulations \citep[e.g.,][]{berlind2003} show that the distribution of
galaxies within a halo is close to Poisson in the
high-occupancy regime, i.e., when $N_{\rm sat} \gg 1$, and (strongly) 
sub-Poissonian in the low-occupancy regime. In order to agree with 
these results, satellite galaxies are assumed to be Poisson
distributed at fixed halo mass. The distinction between central and 
satellite galaxies then automatically accounts for the sub-Poissonian 
behavior of the HOD in the low-occupancy regime \citep{zheng2005}.

The 1- and 2-halo power spectra are
\small
\begin{eqnarray}P_{\rm 1h}(k,z) & = & \int_{\cal{M}}n_{\rm 
halo}(M,z)[2N_{\rm cen}(M)N_{\rm sat}(M) u_{\rm DM}(k,z|M) + \nonumber 
\\
& &  ~~~~~~ N^2 _{\rm sat}(M) u^2 _{\rm DM}(k,z|M) ] dM/n_{\rm 
gal}^{2}(z),\nonumber \\
& & \nonumber \\
  P_{\rm 2h}(k,z) & = & P_{\rm DM}(k,z)\times \nonumber \\ 
  & &  ~~~~~~  \Big{[} \int_{\cal{M}}n_{\rm halo}(M,z)N_{\rm gal}(M,z)\times 
\nonumber \\
  & &   ~~~~~~ b(M,z)u_{\rm DM}(k,z|M)dM\Big{]}^{2}/n_{\rm gal}^2(z). \nonumber 
\\
  \label{eq:P2h}
\normalsize
\end{eqnarray}
The meaning of the symbols here is as follows:
$P_{\rm DM}$ is the linear power spectrum of dark matter, derived 
using the recipes of \citet{eisenstein1998} for the matter transfer 
function; $n_{\rm halo}$ is the halo-mass function 
\citep[see][]{sheth2001}; $b$ is the linear bias parameter; $u_{\rm 
DM}$ is the normalized dark matter halo density profile in Fourier 
space; and $n_{\rm gal}$ is the mean number of galaxies per unit comoving 
volume at redshift $z$,
\small
\begin{equation}
  n_{\rm gal}(z) = \int_{\cal{M}}n_{\rm halo}(M,z)\Bigg{[} 1 + \left( 
\frac{M}{M_1} \right) ^{\alpha}\Bigg{]}dM.
\normalsize
\end{equation}
The expression for the 1-halo term implicitly assumes that the 
distribution of
galaxies traces that of the dark matter, for which we have adopted the 
profile of \citet*[][NFW]{nfw1997}, with the same concentration 
parameter as \citet{bullock2001}.  Since the NFW profile formally 
extends to infinity, it is necessary to artificially truncate the 
distribution at some radius, $r_{\rm cut}$.  Typically, this is chosen 
to be the virial radius of the halo
; however, this may not necessarily be realistic.  We address this by 
first adopting the assumption that 
$r_{\rm cut} = r_{\rm vir}$, and then exploring the consequences of 
relaxing that requirement, so that galaxies are allowed to lie further 
out.  On large scales, where clustering is predominantly linear, 
$u_{\rm DM} \sim 1$, so that the 2-halo power spectrum simplifies to 
$P_{\rm 2h} = b^2_{\rm eff}(z)P_{\rm DM}(k, z)$, where $b_{\rm eff}(z)$ 
is the effective large-scale bias,
\small
\begin{equation}
  b_{\rm eff}(z) = \int_{\cal{M}}n_{\rm halo}(M,z)N_{\rm gal}(M)b(M,z)\ 
dM/n_{\rm gal}(z).
\label{eqn:beff}
\end{equation}
\normalsize
Our model has two free parameters, $M_{\rm min}$ and $\alpha$, which we 
vary through $0 \le \alpha \le 2$ and $10 \le\mathrm{log}(M_{\rm 
min}/{\rm M_{\odot}} )\le 16$, with steps of 0.05 in both $\log M$ and 
$\alpha$.  The best-fit values of the parameters are determined through 
a $\chi^2$ minimization technique by fitting the observed power spectrum at 
each of the three BLAST bands simultaneously. 

Throughout
we assume that both $M_{\rm min}$ and $\alpha$ remain constant in time, 
although in principle they are functions of redshift.  Whether these 
parameters evolve with redshift would be difficult to constrain from 
our data alone; nevertheless, our assumption is consistent with what is 
observed for other classes of high-redshift sources \citep[e.g.,
quasars, see][]{porciani2004}. For each $M_{\rm min}$-$\alpha$ pair, 
the mass-scale $M_1$
is fixed by requiring that at every redshift, $z$, the number density 
of the background sources derived from the halo model formalism matches 
that predicted by the adopted source-population model, i.e.,
\small
\begin{equation}
  \int_{0}^{\infty}\frac{dN}{dS\ dz}(S,z)\ dS = n_{\rm gal}(z)\ 
dV_{\rm c}(z).
\end{equation}
\normalsize
%


\begin{figure*}[t]
  
\hspace{0.0cm}\vspace{0.0cm}\includegraphics[height=12.0cm,width=18.5cm]
  {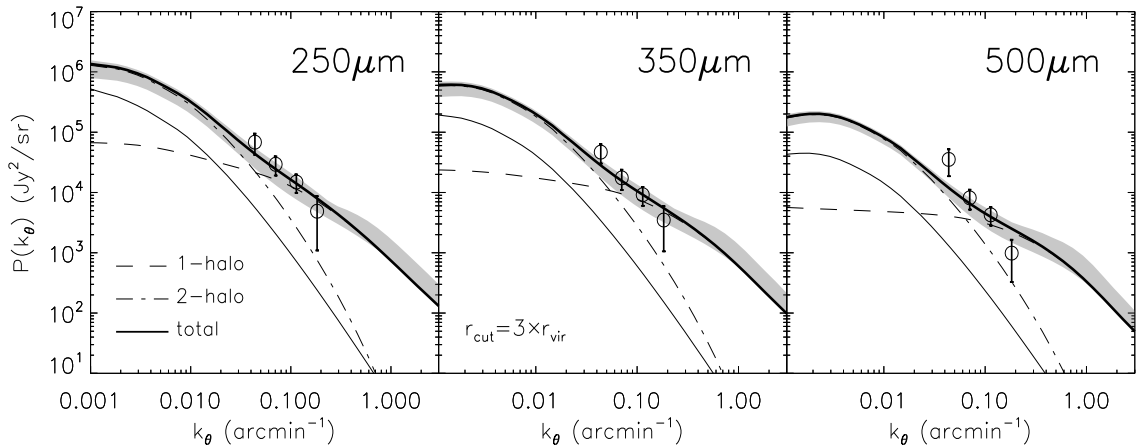}
  \vskip-4.7truecm\caption{Power spectrum of background correlations 
from clustering of extra-galactic sources measured in the BLAST maps 
(circles with error bars) overlaid with the best-fit halo model (thick 
solid line), under the assumption that dark matter halos are NFW 
spheres truncated to $3 \times$ the virial radius, and the galaxies 
within the halo follow the underlying dark matter distribution.
The shaded region shows the 99\% confidence region in the
$M_{\rm min}$--$\alpha$ parameter space. The dashed and the dot-dashed curves 
show the 1- and 2-halo contributions to the power spectrum, 
respectively. For comparison, the power spectrum obtained under the 
assumption that galaxies are unbiased tracers of the underlying dark 
matter distribution, i.e., $P_{\rm 3D} (k, z) = P_{\rm DM} (k, z)$, is 
plotted as a lighter solid line.  According to the model, the BLAST data 
occupy a range of angular scales which should be sensitive to both the linear and 
non-linear clustering terms.} 
 \label{fig:PS_r3}
\end{figure*}
\begin{figure}
  \hspace{0.3cm}\vspace{0.0cm}\includegraphics[height=5.5cm,width=9cm]
  {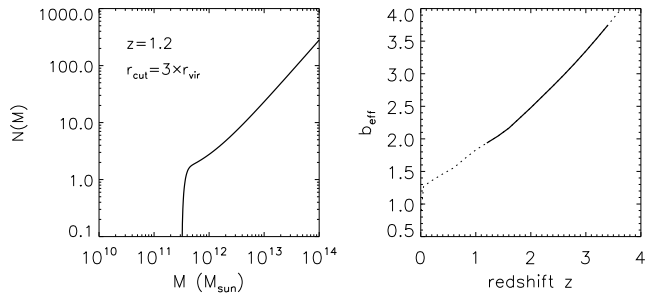}
  \vskip-1.5truecm\caption{{\it Left-hand panel\/}: Halo Occupation
    Number (HON) as a function of the mass of the halo at a 
    representative redshift $z\sim1.2$, assuming the best-fit values
    $M_{\rm min}=10^{11.5}$ M$_{\odot}$ and $\alpha=0.95$.
     {\it Right-hand panel\/}:
    Redshift dependence of the large-scale effective bias,
    resulting from the best fit to the observed power spectrum of
    correlations due to clustering. The redshift
    interval over which 68\% of the signal originates, 
$0.7\lsim z\lsim2.5$, is shown as a thick solid curve.  
}
  \label{fig:HON_r3}
\end{figure}
\begin{figure*}[t]
\hspace{0.0cm}\vspace{0.0cm}\includegraphics[height=12.0cm,width=18.5cm]
  {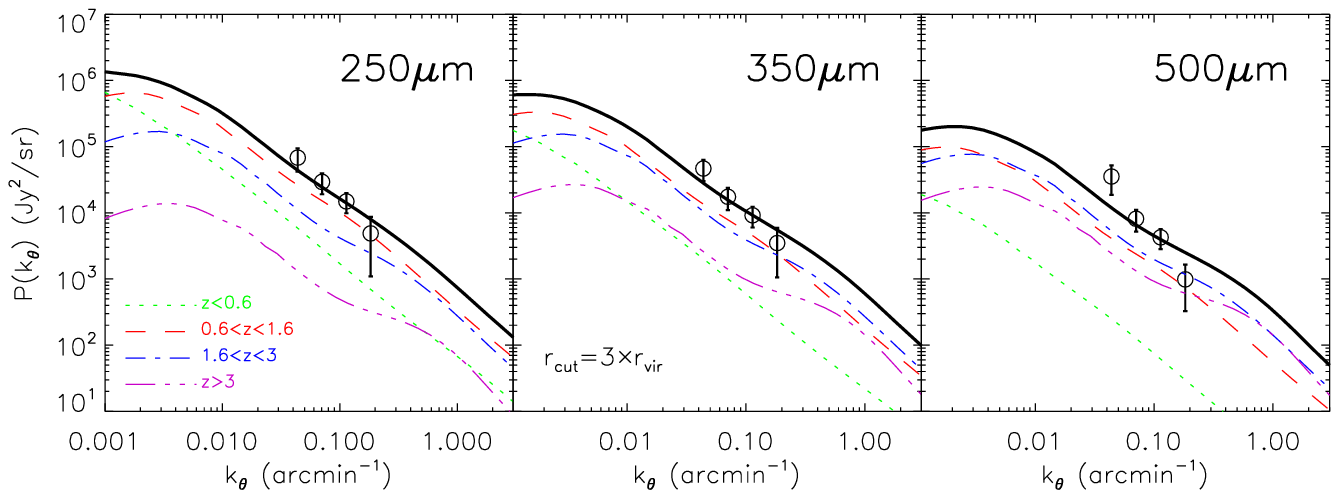}
  \vskip-4.7truecm\caption{Contributions to the total clustering power 
spectrum from sources in increasing redshifts slices.  BLAST 
measurements and the best fit to the halo model (with $r_{\rm cut} = 3 
\times r_{\rm vir}$) are shown as circles with error bars and a solid 
line, respectively.  Overlaid are the contributions from: dotted line 
(green), $0 < z < 0.7$; dashed line (red), $0.7 < z< 1.5$; 
dot-dashed line (blue), $1.5 < z<3.0 $; and triple-dot-dashed line 
(magenta), $z > 3$.  It is clear that at 250\rmicron, the bulk of 
the signal comes from galaxies in the redshift range 0.7--1.5, 
and that the contribution from galaxies in the redshift range 1.5--3.0
increases with increasing wavelength.}
\label{fig:redshift_slices}
\end{figure*}
\section{Halo Model Fits}
\label{sec:results}
Under the assumption that the dark matter halos are described by an NFW 
profile truncated at the virial radius, i.e., where $r_{\rm cut} = 
r_{\rm vir}$, and that galaxies within the halo trace the 
underlying dark matter distribution, 
the best fit to the observed angular power spectrum gives 
$\mathrm{log}(M_{\mathrm{min}}/\rm M_{\odot}) = 11.5_{-0.1} ^{+0.4}$ 
and $\alpha \leq 1.0$, with $\chi^2_{\rm min}=16.3$ with 10 degrees of freedom; i.e., the model is marginally consistent at the 2-$\sigma$ level.   

While this model is not formally ruled out by the data, as shown by a dotted line in Figure~\ref{fig:Pkth_data_vs_model}, it poorly reproduces the shape of the observed power spectrum,
which exhibits a steeper slope. If we were to weight this model 
to fit the large scale power spectrum (e.g., $k \lsim 0.08~ \rm 
arcmin^{-1}$), then it would over-predict the power on small scales.  
Arguing that perhaps the model describes the small scale, 1-halo term 
correctly, but is under-predicting the large-scales, is not a good description 
because: (i) the 2-halo term is less sensitive to the underlying assumptions 
than the 1-halo term \citep[for a discussion see][]{tinker2009}; and (ii) the 
observed small-scale power spectrum is still 
too steep to be accounted for by the shallower 1-halo term.  Another 
possibility is that the redshift distribution of the background 
cumulative flux predicted by the adopted source count model is 
incorrect.  In order to reproduce the observed shape of the 
angular power spectrum, the bulk of the background would have to 
originate from sources at $z < 1$, which is ruled out by 
\citet{devlin2009}, \citet{marsden2009}, and \citet{pascale2009}.  
However, a full investigation of the leeway in changing the redshift
distribution (and degeneracies with other changes to the model) are
beyond the scope of the present study.

In light of this, we explore the possibility that the discrepancy between the 
predicted power spectrum and the observed one is related to the 
modeling of the 1-halo term.  There are two obvious ways to modify the 
shape and normalization of the 1-halo power spectrum.  One is to allow 
the dark matter halos, although still following an NFW profile, to 
be truncated at a scale $r_{\rm cut}>r_{vir}$.  Thus, 
satellite galaxies are distributed over a larger volume.
This idea is not new; \citet{magliocchetti2003} find that in order to 
adequately fit the 1- and 2- halo term to the 2dF Galaxy Redshift 
Survey data-set it is necessary that the galaxies are allowed to reside 
out to 2 times the virial radius.  Furthermore, from semi-analytic 
models \citet{diaferio1999} show that blue (and hence star-forming)
galaxies tend to reside in 
the outskirts of their host halos, while red galaxies are found 
closer to the halo center.  

The second possibility is that the distribution of galaxies within the 
halos does not follow that of the underlying dark matter. For example, 
a power-law distribution $\rho(r) \propto r^{- \gamma}$ with $\gamma < 
2$ would make the 1-halo angular power spectrum steeper than that 
predicted by an NFW profile.  Similar arguments have been made by \citet{watson2009}, who found that by allowing the inner slope of the density profile to vary, they fit the small-scale clustering of luminous red galaxies quite well.

Here we only explore the first possibility, examining $r_{\rm cut} = 1, 2, 3,$ 
and $4 \times r_{\rm vir}$.  The results are shown in 
Figure~\ref{fig:Pkth_data_vs_model}, and summarized in 
Table~\ref{tab:chi2}.  The best-fit model angular power spectrum for 
the case $r_{\rm cut} = 3 \times r_{\rm vir}$ is shown in 
Figure~\ref{fig:PS_r3}, while the corresponding HON and large-scale 
effective bias are shown in Figure~\ref{fig:HON_r3}.  Note that while 
the value of the reduced $\chi^2_{\rm min}$ approaches unity for increasing 
values of $r_{\rm cut}$, the best-fit values of $M_{\rm min}$ are 
negligibly affected by the changes, while $\alpha$ only marginally 
increases with increasing $r_{\rm cut}$. 
The effective mass of the halo, $M_{\rm eff}$, is the weighted mean over the halo-mass distribution:
\small
\begin{equation}
  M_{\rm eff}(z) = \int_{\cal{M}}n_{\rm halo}(M,z)N(M,z)MdM/n_{\rm 
gal}(z).
\normalsize
\end{equation}
The results for the large-scale effective bias $b$, and mass log($M/\rm M_{\odot}$), 
at the respective medians of the redshift distributions of the sources contributing to the background in each band (see Figure~\ref{fig:rds}) are
$2.2\pm0.2$, $2.4\pm0.2$, and $2.6\pm0.2$, and $12.9\pm 0.3$, $12.8\pm 0.2$, and $12.7\pm 0.2$, at 250, 350 and 500\rmicron, respectively.  They are minimally affected by the change in $r_{\rm cut}$, changing by less than 8\% for the bias, and 3\% for the mass, over the full range of $r_{\rm cut}$ expored.

Figure~\ref{fig:redshift_slices} shows the contribution to the total 
clustering power spectrum from sources in different redshift slices, within
the assumed source population model.  
As expected, the power is dominated by the contribution from sources in 
the range $0.7 < z< 1.5$, with an increasing contribution from 
sources at $z=1.5-3.0$ for increasing wavelengths, as is expected from Figure~\ref{fig:rds}.  
This is consistent with the findings of \citet{devlin2009}; 
\citet{marsden2009}; and \citet{pascale2009}, who through stacking show 
that of the sources making up the CIB, the fraction at $z > 1.2$ 
increases from 40\% at 250\rmicron, to 50\% and 60\% at 350, and 500\rmicron. 

Finally, we can use the model to interpret the clustering in terms of a 3D spatial correlation length, $\rm r_0$.  To do that, we Fourier transform the best-fit power spectrum to find the spatial correlation function, $\xi(r)$, from which $\rm r_0$ is then simply the linear comoving scale at which the correlation function equals 1 at each redshift.  This is a model-dependent approach to estimating $\rm r_0$, and as such should be considered an approximation.  The more typical approach, which involves finding the angular correlation length, the redshift distribution, and deprojecting the signal by inverting the Limber equation, would result in very large uncertainties.  The model-dependent $\rm r_0$ is only mildly sensitive to the choice of $r_{\rm cut}$, varying by 10\% over the full range.
The model-dependent values for $\rm r_0$ are illustrated in Figure~\ref{fig:r_plot} as a solid line with a shaded area representing 3-$\sigma$ uncertainties.  The three BLAST points are plotted at the flux-weighted median redshifts of the unique redshift distributions probed by the three bands (i.e., the distributions shown in Figure~\ref{fig:dSdz_LagacheModel}), and should not be interpreted as the locations of all the sources contributing power in those bands.  At these effective redshifts, (i.e., $z = 1.60, 1.86$ and $2.15$), we find $\rm r_0 = 4.9$, 5.0, and $5.2~\pm 0.6. ~ h^{-1}~ \rm Mpc$ at 250, 350 and 500\rmicron, respectively.\\  

\section{Discussion}
\subsection{Comparison with other observations}
Comparisons with other measurements of clustering must be made and interpreted with care because not everyone uses the same definition of bias, the same parameterization for the halo model, etc.  
Nevertheless it is interesting to put our measurements into the context of the large body of literature on the clustering of galaxies selected in different ways, in order to understand how they might be related.
\begin{figure*}[t]
   \begin{center}
\hspace{0.0cm}\vspace{0.0cm}\includegraphics[height=8.7cm,width=13.9cm]
  {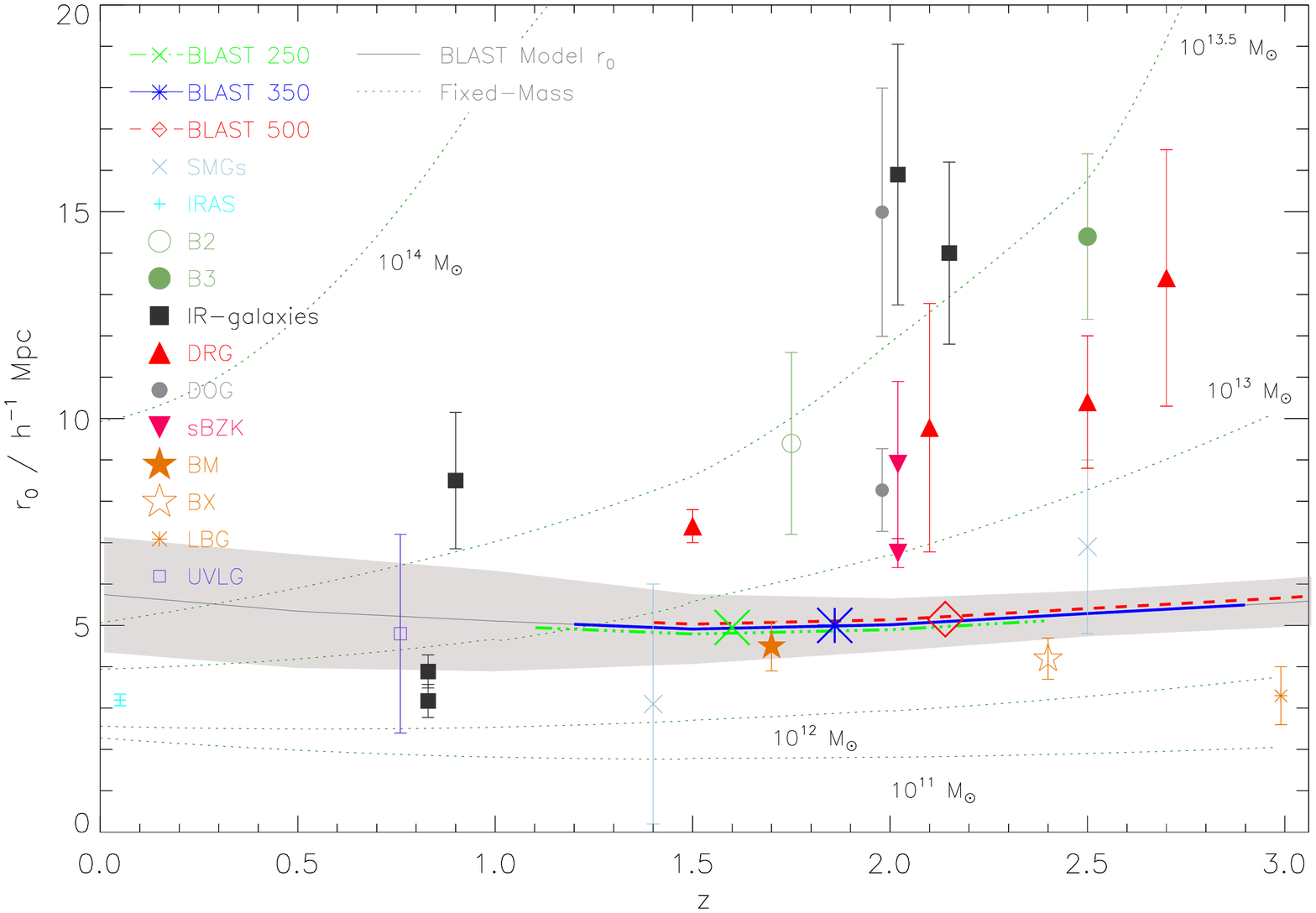}
  \vskip-.1truecm\caption{Comoving correlation length vs.\ redshift for 
star-forming galaxies selected with a variety of techniques.  Other 
data taken from: {IRAS\/} -- \citet{saunders1992}; SMG -- \citet{webb2003a, 
blain2004}; B2 and B3 -- \citet{farrah2006}; IR -- 
\citet{magliocchetti2007, magliocchetti2008a, gilli2007}; 
DRG -- \citet{grazian2006, quadri2008}; DOG -- \citet{brodwin2008}; BzK -- 
\citet{blanc2008, hartley2008}; BM and BX -- \citet{adelberger2005}; LBG 
-- \citet{giav1998, adelberger2005}; 
and 
UVLG -- \citet{basu2009}
.  The dotted fixed mass lines show the predicted clustering 
lengths of halos of a given mass at a given redshift, found with the \emph{Millennium Simulation\/} \citep{springel2005}.
The model-dependent BLAST values for $\rm r_0$ are shown as a solid line with a shaded area representing 3-$\sigma$ uncertainties.  The three BLAST points are plotted at the median redshifts of the distributions from which the signal originates (i.e., the distributions shown in Figure~\ref{fig:rds}).  The ranges from which 90\% of the power originates are illustrated as corresponding colored lines.  
In the context of the model, the clustering strength of BLAST galaxies is compatible with the \citet{webb2003a} and \citet{blain2004} estimates for clustering of submillimeter galaxies, but less strong than the that of other resolved populations of galaxies.  
}  
  \label{fig:r_plot}
 \end{center}
\end{figure*}

In Figure~\ref{fig:r_plot} we compare the correlation lengths vs. redshift of star-forming populations selected using a wide variety of techniques.  It should be noted that in some cases these techniques select overlapping populations 
\citep[see][for a nice discussion on the overlap between color selected samples]{reddy2005}.  This list is by no means exhaustive, and is meant only to be an illustration.  The reported values of ${\rm r_0}$ were converted assuming a fixed slope $\gamma = 1.8$, such that ${\rm r_{0, 1.8} = (r_{0, \gamma})^{-\gamma / 1.8}}$.  
The BLAST best-fit model estimates are shown as a line and shaded 3-$\sigma$ confidence region.   The three BLAST points are located at the median locations of their respective redshift distributions (see Figure~\ref{fig:rds}). 
In addition, the simulated clustering lengths of dark matter halos of given mass and redshift are measured from the \emph{Millennium Simulation\/}\footnote[5]{Note that Millenium Simulation cosmology is $\sigma_8 = 0.9$ and $\Omega_{\rm M}=0.25$, which has a minimal impact on the prediction.  Catalogues can be found at http://www.mpa-garching.mpg.de/Millennium} \citep{springel2005}, by fitting a single power-law with slope of -1.8 to the correlation function, and are shown as dotted lines.

It is immediately clear that the galaxies which make up the background are not as strongly clustered as the more luminous sets of resolved sources (with the exception of those identified by their Lyman break, i.e., BM, BX and LBG).  Furthermore, the strength of the clustering increases with increasing luminosity \citep[e.g.,][]{gilli2007,brodwin2008}.  Since each of the techniques used to select the populations of galaxies that lie above the BLAST lines has an IR component, it is tempting to conclude that all of these populations contribute to the total submillimeter CIB.  The relative contribution of each of these populations could be explored through stacking, as \citet{marsden2009} have done for BzKs.  They found that although the BzKs make up about a quarter of the sources which completely resolve the CIB, they contribute $\sim 32\%$, $\sim 34\%$, and $\sim 42\%$, at 250, 350, and 500\rmicron, to the total BLAST intensity.  This indication that the resolved sources pick out parts of the background highlights the complementary nature of the clustering measurements of the CIB and resolved sources in forming a complete picture of the environments of star-forming galaxies.

Although the correlation length of the background galaxies appears to change very little with redshift, the bias is a strong function of redshift (see Figure~\ref{fig:HON_r3}).  While both the bias and the correlation length are indicators of galaxy clustering, their behavior is not in contradiction because the clustering strength of the host dark matter halos is rapidly increasing with decreasing redshift as well, thus in this sense, it is the bias that is a more telling description of how star formation relates to structure formation.
This strong evolution of the bias is confirmed by \citet{lagache2007}, who found a redshift-independent bias parameter, $b\sim2.4$, for the sources which make up the CIB at 160\rmicron\footnote[6]{Note that the value $b=1.7$ reported in \citet{lagache2007} is for $\sigma_{8}=1.1$, and not $\sigma_{8}=0.8$, as quoted in the text.  The correct value of the measured bias parameter is $2.4\pm0.2$ (Lagache, private communication).}, at $z \sim 1$.  Our earlier result (see Section~\ref{sec:linear}), for galaxies which lie at higher redshifts, was $b\simeq 4$.  Thus, in the scenario of a strongly evolving bias parameter, from anti-biased in the local Universe, to highly biased at $z \sim 1$ and beyond \citep[e.g.,][]{sheth2006, elbaz2007}, our result is consistent.
\subsection{Clustering in Context}
\citet{lefloch2005} show that IR-luminous galaxies (LIRGs and ULIRGS) represent $\sim 70\%$ of the IR energy density, and are responsible for most of the star formation, at $z \sim 0.5$--1.0 and beyond.  Stacking work \citep[e.g.,][]{marsden2009} shows that most of the objects responsible for the CIB are fainter than the flux density limit of the BLAST catalogs \citep{devlin2009}, corresponding to LIRG-like luminosities for those sources.  Therefore, by identifying the locations of active star formation, we are also identifying the locations of the formation of the majority of stars in the present day Universe.  With this in mind we ask: where are active star-forming galaxies preferentially found through cosmic time?
From Figure~\ref{fig:r_plot}, it appears that the most active star-forming galaxies are found occupying halos whose mass increases from roughly $10^{12}$ to $10^{13.5}~ \rm M_{\odot}$ over the redshift range $z = 0$ to 2, after which it appears to remain roughly constant.  This appears to be consistent with \emph{downsizing\/} \citep{cowie1996}, the scenario in which the sites of most active star formation shifts to ever larger galaxies at higher redshifts. 
However, according to the model, which is only constrained for redshifts greater than $\sim 1.1$, the galaxies which make up the background do not exhibit the same trend.  While the bias is still a strong increasing function of redshift, the clustering strength of the host halos remains roughly constant, corresponding to typical host halos that become slighly smaller.  Since this is very much a model-dependent claim, it may be indicitave of a flaw in the model (perhaps assuming $M_{\rm min}$--$\alpha$ remain constant with $z$ is incorrect); future studies should clarify this picture.

A striking feature is the sharp cut-off at $M \ge 10^{13.5}~ \rm M_{\odot}$, which appears to hold out to $z \sim 2.5$.  As was pointed out by  \citet{brodwin2008}, this appears to be inconsistent with models which claim that star formation should be quenched in halos with masses greater than $10^{12}~ \rm M_{\odot}$ due to shock heating \citep{birnboim2003, dekel2006, cattaneo2008}.  
\citet{dekel2009} attempt to resolve this dilemma with a model where cold streams penetrate the shock-heated media.  On the other hand, if the shock radius roughly follows the virial radius \citep{birnboim2003}, then finding satellites actively forming stars outside of the shock-heated volume would satisfy both the model and the observations.

\section{Conclusions}
\label{sec:conclusions}
We report the detection of correlations in excess of Poisson noise in 
the CIB, over scales of approximately 5--25 arcmin, with BLAST at 250, 350, and 500\rmicron, at a level with 
respect to the CIB of $\Delta I/ I = 15.1 \pm 1.7\%$.

The CIB is made almost entirely out of individual sources distributed over a wide range of redshifts.  
We find that within the
context of a reasonable model for the source population, the signal
originates from galaxies in the redshift ranges ${\rm 1.3 \le z \le 2.2,~ 1.5 \le z \le 2.7,~ and~ 1.7 \le z \le 3.2}$, with median redshifts $z = 1.61, 1.88$ and $2.42$, at 250, 350 and 500 \rmicron, respectively.  Fitting to the linear theory power
spectrum, we find that the BLAST galaxies responsible for the CIB
fluctuations have a bias parameter, $b = 3.8\pm 0.6, 3.9\pm 0.6$ and $4.4 \pm 0.7$.

We further interpret our results in terms of the halo model.  We find that
the simplest prescription does not fit very well.  One way to improve the
fit is to increase the radius at which we artificially truncate dark matter halos to well outside the
virial radius.  This may imply that the star-forming galaxies that we are
seeing at $z\sim1$ are preferentially found in the outskirts of groups and clusters.
This is consistent with related phenomena that have been observed at other
wavelengths \citep{magliocchetti2004, marcillac2007}, as well as in
simulations \citep{diaferio1999}.

For a HOD with \lq satellite\rq\ galaxies occupying halos out as far as $r_{\rm cut}=3r_{\rm vir}$, we find parameters $\log(M_{\rm min}/{\rm M}_\odot =11.5^{0.4}_{-0.1}$, and $\alpha=1.1^{+0.8}_{-0.1}$, resulting in effective biases $b_{\rm eff} = 2.2 \pm 0.2, 2.4 \pm 0.2 $, and $2.6 \pm 0.2 $, and effective masses $\mathrm{log}(M_{\mathrm{\rm eff}}/\rm M_{\odot}) 
= 12.9\pm 0.3$, $12.8\pm 0.2$, and $12.7\pm 0.2$ at 250, 350 and 500\rmicron, corresponding to spatial correlation lengths of $\rm r_0 = 4.9, 5.0,$ and $5.2~\pm 0.7 ~ h^{-1}~ \rm Mpc$, respectively.

In the context of the model, we see that star formation is highly 
biased at $z\gsim 1$, unlike in the local Universe, where analogous 
galaxy populations, such as \emph{IRAS\/} galaxies, are found to be 
mildly anti-biased \citep[e.g., $M_{\rm eff}\lsim10^{11}$ M$_{\odot}$ and 
$b = 0.86$,][]{saunders1992}.  

We find relatively small values for $\theta_0$,
further confirming that the sources which make up the CIB are distributed over a wide range of redshifts, which has implications for planning future submillimeter clustering measurements.  For example, as \citet{knox2001} have argued, to match the precision of the bias 
measurement made from correlations in the background by using discrete 
sources will be very challenging.  To achieve $\sim 5$\% accuracy would 
effectively require thousands of sources with exact redshifts, over
tens of square degrees.  When redshifts are only approximately known, 
this increases to hundreds of thousands of sources over
hundreds of square degrees, which is only possible with instruments 
whose beams are smaller than 5 arcsec.  Thus, while measuring the 
clustering from resolved sources has numerous advantages -- for 
example studying the clustering properties of a subset of galaxies -- 
accurately measuring the large scale bias is best achieved through 
correlation analysis of the background fluctuations.  This will be a 
focus of future studies with BLAST, as well as with the
\emph{Herschel\/} and \emph{Planck\/} satellites, and SCUBA-2. 

\newpage
\begin{acknowledgments}
We acknowledge the support of NASA through grant numbers NAG5-12785,
NAG5-13301, and NNGO-6GI11G, the NSF Office of Polar Programs, the 
Canadian
Space Agency, the Natural Sciences and Engineering Research Council 
(NSERC) of
Canada, and the UK Science and Technology Facilities Council (STFC).  
CBN acknowledges support from the Canadian Institute for Advanced 
Research.  
We thank the anonymous referee for her/his excellent comments, which 
improved this paper greatly, and Guilaine Lagache, Ravi Sheth, and Adam 
Muzzin for useful discussions and helpful advice.
MPV extends his warm thanks to Olivier Dor\'{e} for his help and 
infinite patience, and Silvia Bonoli for kindly providing simulated 
clustering lengths of dark matter halos.
\end{acknowledgments}

\bibliographystyle{apj}
\bibliography{apj-jour,refs}

\end{document}